\documentclass[%
 amsmath,amssymb,
 aps,pra
]{revtex4-1}

\usepackage{graphicx}
\usepackage{dcolumn}
\usepackage{bm}
\usepackage{hyperref}

\usepackage{braket} 

\usepackage{siunitx}

\begin{document}


\title{Time-reversed and coherently-enhanced memory:\\ A single-mode quantum atom-optic memory without a cavity.}

\author{J.L. Everett}
\author{P. Vernaz-Gris}
\author{G.T. Campbell}
\author{A.D. Tranter}
\author{K.V Paul}
\author{A.C. Leung}
\author{P.K. Lam}
\author{B.C. Buchler}
\affiliation{Centre for Quantum Computation and Communication Technology, Department of Quantum Science, The Australian National University, Canberra, Australia}

\date{\today}

\begin{abstract}
The efficiency of an ensemble-based optical quantum memory depends critically on the strength of the atom-light coupling. An optical cavity is an effective method to enhance atom-light coupling strength, with the drawback that cavities can be difficult to integrate into a memory setup. In this work we show coherent enhancement of atom-light coupling via an interference effect. The light to be absorbed into the atomic ensemble is split and used to drive the atoms from opposite ends of the ensemble. We compare this method theoretically to a cavity enhanced scheme and present experimental results for our coherent enhancement in cold rubidium-87 atoms that show an efficiency of $72\pm5\%$ and a storage lifetime of $110\pm 10 \si{\micro\second}$.

\end{abstract}

\maketitle
\section{Introduction}
Atom-optic memories operate by the coherent reversible absorption of light into atomic excitations \cite{gorshkov_photon_2007,nunn_multimode_2008,moiseev_optical_2011}.
Quantum memories in high optical depth ensembles can theoretically reach near-unity memory efficiency, which is a crucial figure of merit for their application in future quantum information and computation networks \cite{briegel_quantum_1998,kok_linear_2007,sangouard_quantum_2011}.

Restricting our consideration to a single spatial mode, a distinction can be drawn between quantum memories that aim to store a single temporal mode and memories that aim to store all modes within a given bandwidth, whether separated in the temporal or frequency domains. A multi-mode memory stores distinguishing information about the modes as internal states of the memory such as longitudinal position \cite{fleischhauer_quantum_2002,geng_electromagnetically_2014}, phase including inhomogeneous broadening \cite{moiseev_complete_2001,alexander_photon_2006,mishina_spectral_2007,sangouard_analysis_2007,afzelius_multimode_2009}, or both \cite{hetet_photon_2008,cho_highly_2016}. A single-mode memory optimises the storage of a single time-frequency mode and there is no mapping of the mode to internal variation of the memory \cite{gorshkov_photon_2007}. A single-mode memory is typically implemented by placing the absorbing medium in a cavity \cite{bao_efficient_2012,jobez_cavity-enhanced_2014,bimbard_homodyne_2014,saunders_cavity-enhanced_2016}. The input optical field traverses the memory many times while being absorbed or re-emitted and so interacts uniformly along the memory. In a multi-mode memory, the interaction must instead be designed so that the mapping into the internal state can be reversed, accurately reproducing the input upon recall. 


Recently,  spatially distributed reflection or photonic bandgap for light in  atomic media has been investigated for its potential to enhance nonlinear optical interactions \cite{hafezi_photonic_2011,shahmoon_highly_2016,albrecht_changing_2017,lahad_induced_2017,iakoupov_controlled-phase_2018,murray_coherent_2017,asenjo-garcia_exponential_2017,song_photon_2018}. The bandgap can result from a Bragg reflection from spatially ordered atoms \cite{iakoupov_dispersion_2016,corzo_large_2016,song_photon_2018}, and spatially modulated driving of the atom-light interaction \cite{arkhipkin_raman-induced_2014,iakoupov_dispersion_2016,liu_asymmetric_2017}. In a disordered medium the bandgap can also be generated by a multi-wave mixing process with imposed phase-matching \cite{zimmer_coherent_2006,everett_dynamical_2016,campbell_direct_2017}. We use this second process to implement a single-mode optical memory, where the distributed reflection results in uniform driving of the ensemble. The reflection is generated by coupling of the light into and out of a coherent atomic excitation (a \textit{spinwave}) with a pair of counter-propagating, phase-matched optical control fields generating stimulated Raman scattering. This has similar theoretical efficiency limitations to a memory in a cavity \cite{gorshkov_photon_2007}, but is more accessible to free-space optical fields. 

\section{Theory}
We create a single-mode memory by splitting an input optical pulse in two and sending it to opposite ends of an atomic ensemble. The two inputs are far-detuned from an atomic resonance and Raman interaction generated by two separate control fields couples them to the memory. With appropriate phase-matching and detuning of the fields, the total driving of the ensemble is uniform along the length of the memory. The input fields can be efficiently absorbed in the memory by choosing the appropriate time profile for the control fields, so that the output fields interfere destructively with the input. The achievement of a single-mode memory dependent on the coherent interaction of multiple driving fields gives the name, a `time-reversed and coherently-enhanced' (TRACE) memory.

3-level ensembles are a well understood platform for optical memories. A transition at optical frequency allows efficient absorption of the input optical field, and the excited state is coupled to a long-lived state by an optical control field, allowing long memory lifetime and control over the coupling into and out of the memory. The ensemble allows a quantum memory due to the coherent collective character of the absorption and re-emission. 

A coherent memory is also sensitive to phase, which in a 3-level ensemble is the local relative phase between the control field and the input field. A spatially extended memory generally means a different relative phase at different locations if the control field has different frequency, dispersion or direction to the input. It is important to match this phase at retrieval for efficient and high fidelity recall, which generally occurs automatically for retrieval into the same optical mode as storage. However, some memory schemes require backward retrieval to achieve optimal efficiency and phase-matching is then a consideration. This is often achieved by the spatial arrangement of the control fields.

In addition to storing and retrieving in different directions, the TRACE memory adds the complication that these different directions are coupled simultaneously during storage and retrieval. The two control fields each drive the absorption/reemission of one of the probe fields and, in addition to the phase-matching requirement, the two interactions must be matched in global phase to ensure the memory excitation generated by each interferes constructively. The equations below assume phase matching and equal global phases. We did not match global phase experimentally but we discuss methods for doing so below.

\begin{figure}\centering
\includegraphics[width=0.5\textwidth]{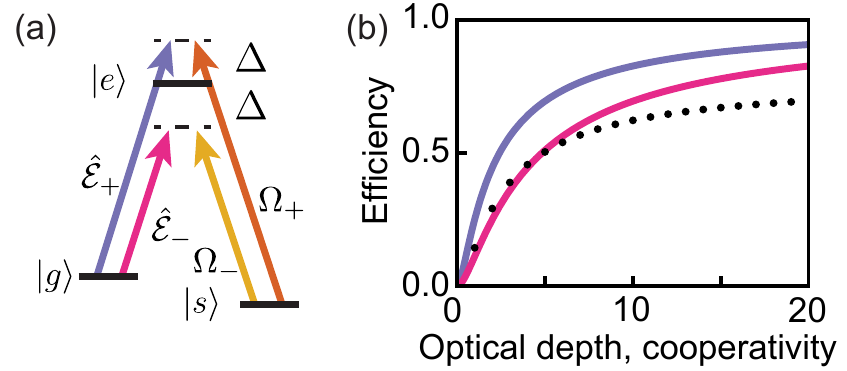}
\caption{(a) Simplified atomic level scheme for the memory interaction. (b) Dependence of efficiency in a cavity storage and retrieval on cooperativity (blue), dependence of TRACE storage and retrieval efficiency on optical depth (magenta). Approximate optimal efficiencies for a free-space Raman memory (black dots).\label{fig:odcurve}}
\end{figure}

We can describe the evolution of a spinwave over a normalised spatial coordinate $z~\epsilon~\left[0,1\right]$ driven by two phase-matched classical optical control fields, each with Rabi frequency $\Omega(t)$, and weak input optical fields $\hat{\mathcal{E}}_+(z=0,t)$ and $\hat{\mathcal{E}}_-(z=1,t)$. The subscript $\pm$ refers to fields travelling along $z$ or $-z$ respectively. The spinwave $\hat{S}$ describes the coherent superposition of atoms in the ground state $\ket{g}$ and a hyperfine state $\ket{s}$ (as shown in Fig. \ref{fig:odcurve} (a)) and evolves according to
\begin{align}
(\partial_t+\gamma)\hat{S}(z,t) &= i\sqrt{d}\Gamma\frac{\Omega(t)}{\Delta}\left(\hat{\mathcal{E}}_+(z,t)+\hat{\mathcal{E}}_-(z,t)\right)\label{eq:spinwave}\\
\partial_z\hat{\mathcal{E}}_\pm(z,t) &= \pm i\sqrt{d}\frac{\Omega(t)}{\Delta}\hat{S}(z,t)\label{eq:efield}
 \end{align}
where $d$ is the optical depth of the atomic ensemble, $\Gamma$ is half the natural linewidth of the transition $\ket{e}\rightarrow\ket{g}$, and $\Delta$ is the single-photon detuning from this transition. The optical energies relative to the transition energies are shown in Fig. \ref{fig:odcurve} (a). The assumption $\Delta \gg \Gamma$ allows the effect of the excited state to be ignored, excepting for its contribution to the decay of the spinwave, $\gamma$. Eq.~\eqref{eq:memory} is obtained by inserting Eqs.~\eqref{eq:efield} into \eqref{eq:spinwave}.

\begin{align}
\partial_t\hat{S}(t) &= -d\Gamma\frac{\Omega(t)^2}{\Delta^2}\int_0^1{\hat{S}(z',t)\mathrm{d}z'}\nonumber\\ &+ i\sqrt{d}\Gamma\frac{\Omega(t)}{\Delta}\left(\hat{\mathcal{E}}_+(z=0,t)
+\hat{\mathcal{E}}_-(z=1,t)\right)\label{eq:memory}
\end{align}

Equation \eqref{eq:memory} shows the time evolution of the spinwave is independent of position, indicating a uniform driving of the memory. 

Efficient memory operation requires complete absorption of the input fields, which we write as $\hat{\mathcal{E}}_{IN}(t)=2\hat{\mathcal{E}}_+(z=0,t)=2\hat{\mathcal{E}}_-(z=1,t)$ 
The fields exiting the memory as output are $\hat{\mathcal{E}}_{OUT}(t)=2\hat{\mathcal{E}}_+(z=1,t)=2\hat{\mathcal{E}}_-(z=0,t)$ 
The optical field $\hat{\mathcal{E}}_{OUT}$ depends on the stored field as well as the coupling parameters, and an expression for the output field can be found by assuming the previous field was completely absorbed:
\begin{align}
\hat{\mathcal{E}}_{OUT}(t)&= \hat{\mathcal{E}}_{IN}(t)+i\sqrt{d}\frac{\Omega}{\Delta}\int_0^1{\hat{S}\mathrm{d}z'} \\
&= \hat{\mathcal{E}}_{IN}(t)-2d\Gamma\frac{\Omega^2(t)}{\Delta^2}\int_{-\infty}^t{|\hat{\mathcal{E}}_{IN}(t')|\mathrm{d}t'}
\end{align}

By setting $\hat{\mathcal{E}}_{OUT}=0$ we can solve for the coupling field time profile required to store a particular temporal mode of the input:
\begin{align}
\Omega(t)&=\frac{\Delta}{\sqrt{2d\Gamma}}\frac{\hat{\mathcal{E}}_{IN}(t)}{\sqrt{\int_{-\infty}^t{|\hat{\mathcal{E}}_{IN}(t')|^2\mathrm{d}t'}}}
\end{align}

For example, constant coupling fields can be used to store an exponentially rising input described by $\hat{\mathcal{E}}_{IN}=C_1\exp(d\Gamma(\Omega/\Delta)^2t)$. The control fields can be switched off to prevent re-emission at the end of the input, and switched on at a later time to recall the input with an exponentially decaying time profile.

\begin{figure}[htbp!]
\centering
\includegraphics[width=0.5\textwidth]{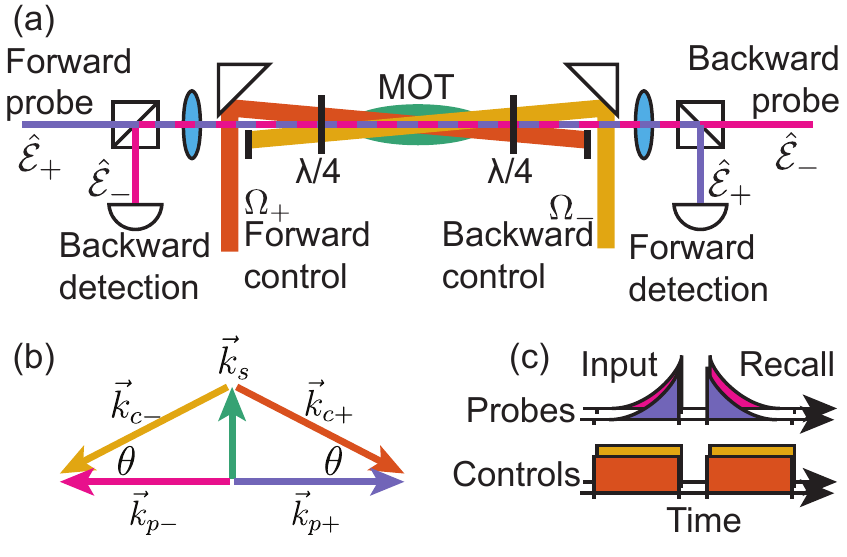}
\caption{(a) Experimental layout. The two counterpropagating probe fields $\hat{\mathcal{E}}_\pm$ are combined in the same spatial mode and then separated for detection after travelling through the ensemble by non-polarising beamsplitters. The control fields $\Omega_\pm$ are overlapped with the probe fields within the ensemble using wedge mirrors. (b) At the phase-matching condition $\theta=6$ mrad, the momentum differences of each pair control-probe pair are equal, $\vec{k}_{p+}-\vec{k}_{c+} = \vec{k}_{p-}-\vec{k}_{c-}$ and each generate a spinwave with identical momentum $\vec{k}_s$ (c) Example timing sequence for probe and control fields for storage of probes. \label{fig:layout}}
\end{figure}

These equations \eqref{eq:spinwave}, \eqref{eq:efield}, and \eqref{eq:memory} assume spatial phase-matching of the inputs. This is the assumption that the phases of the spinwave and the inputs do not vary over the length of the memory in the chosen frame. The derivation of the interaction of counter-propagating fields  with three-level atoms and the resulting requirements for spatial phase-matching appear in the supplementary material for a previous work \cite{everett_dynamical_2016}. The experimental geometry of the fields required for spatial phase-matching is shown in Fig. \ref{fig:layout} (a) and (b).

We have shown the scheme acts as a single-mode memory, and are now interested in how it compares to other single-mode memories, as well as other free-space memories. Memories relying on a control field have been shown to have an optimal efficiency that is dependent on optical depth, due to the branching between absorption into the memory and spontaneous decay from the excited state \cite{gorshkov_universal_2007}. Cavity single-mode memories have the highest optimal efficiency; they make the best use of the optical depth due to a uniform interaction. Efficiency of storage followed by retrieval approaches $1-2/C$ for $C\gg 1$, where $C$, or cooperativity, is the optical depth multiplied by the average number of trips the input makes in the cavity \cite{gorshkov_photon_2007}.
 
A free-space memory has a lower optimal efficiency. The requirement that the input be absorbed in a single traversal of the memory means the input field and thus the driving varies across the memory. This limits efficiency to $1-5.8/d$ for $d\gg 1$ \cite{gorshkov_photon_2007-1}.

In the TRACE scheme the sum of the two input fields, each absorbed in a single traversal, results in uniform driving at the cost of adding an additional decay pathway. TRACE has a higher optimal efficiency than other free-space memories but has additional losses compared to single-mode memories with a single coupling field. In Appendix \ref{appendixeffic} we follow the method of Gorshkov et al. \cite{gorshkov_photon_2007}, adding an additional excited state coupling to show the efficiency of storage followed by recall is $1-4/d$ for $d\gg 1$. Fig. \ref{fig:odcurve} (b) compares the efficiencies of these schemes.

\section{Experimental setup}
We implemented this novel memory protocol in a dense ensemble of rubidium-87 atoms, prepared in the elongated magneto-optical trap described in previous works \cite{sparkes_gradient_2013,cho_highly_2016,tranter_multiparameter_2018}. The atomic levels used in the memory protocols were the two hyperfine ground states that we define as $\ket{g}=\ket{5S_{1/2}, F=2, m_F=+2}$ and $\ket{s}=\ket{5S_{1/2}, F=1,m_F=0}$, and the excited state $\ket{e}=\ket{5P_{1/2}, F'=1, m_{F'}=+1}$. We characterised the storage of a superposition of controllable-waveform probes detuned above and below the $\ket{g}\rightarrow\ket{e}$ transition by 160 or 230 MHz  ($\approx 28$ or $40\Gamma$ - in separate experiments). The blue-detuned light was sent in the forward direction and the red-detuned light in the backwards direction in the same spatial mode. The amplitude of the electric field of this probe in the forward and backward directions are denoted $\mathcal{E}_+$ and $\mathcal{E}_-$ respectively, as shown in Fig. \ref{fig:odcurve} (a). A pair of control laser fields in the forward and backward directions were similarly detuned by 160 or 230 MHz from the $\ket{s}\rightarrow\ket{e}$ transition to each address a two-photon transition to the $\ket{s}$. These are denoted $\Omega_+$, as shown in Fig. \ref{fig:odcurve}(a). For the 6.83 GHz hyperfine splitting of rubidium-87, a phase-matching angle of $ \theta_\pm=\pm 6$ mrad was introduced between  $\mathcal{E}_\pm$ and $\Omega_\pm$, which resulted in forward and backward probe-control pairs addressing the spinwave of the same momentum $\vec{k_s}=\vec{k_p}-\vec{k_c}$ (momentum is equivalent to the spatial variation of the phase).

High optical depth and low atomic temperature were obtained by compressing the atomic cloud and applying optical molasses and polarisation gradient cooling techniques.
The degeneracy of the Zeeman sub-levels was lifted by applying a uniform $2.1$ G magnetic field along the optical axis. The atoms were transferred to the $m_F=+2$ state by applying a $\sigma_+$-polarised optical pump resonant between $\ket{5S_{1/2}, F=2}\rightarrow\ket{5P_{3/2}, F=3}$, along with a repump field $\ket{5S_{1/2}, F=1}\rightarrow\ket{5P_{3/2}, F=2}$, for a duration of  $400\ \si{\micro\second}$ after the end of the polarisation gradient cooling phase.
To mitigate the effect of eddy currents in nearby conductors due to the shut-off of the magnetic trapping, the memory sequence began $1\ \si{\milli\second}$ after the end of the optical pumping.
To characterise the optical depth on the probe transition, the frequency of the probe was swept over $6$ MHz to address the two-photon resonances for the $m_F=-1,0,+1$ Zeeman sub-levels of the ground state. The Rabi frequency of the control was calibrated by measuring the ac-Stark shifts of the Zeeman sub-levels for various control field powers. This allowed the characterisation of the optical depth on the  $\ket{g}\rightarrow\ket{e}$  transition by studying the Raman transition.  Optical depths of $d=500\pm100$ were obtained.

\begin{figure}[h!]
\centering
\includegraphics[width=0.5\textwidth]{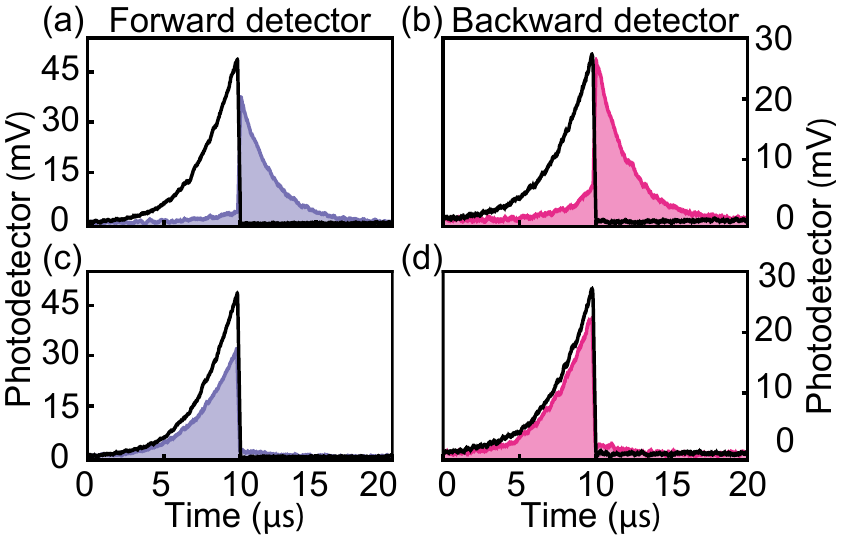}
\caption{Raw data for storage experiments. (a,b) an efficient storage due to constructive interference into the spinwave. (c,d) an inefficient storage due to destructive interference. Black lines are the sent pulses, while filled colours are the transmitted and recalled pulses. \label{fig:traces}}
\end{figure}
\section{Results}
\subsection{Memory efficiency}
Memory efficiency was measured by sending and immediately retrieving 2000 rising exponential pulses. The decay time of the memory was also measured by delaying the recall of the pulses as in Fig. \ref{fig:layout}(c). In this experiment, the two inputs interfered with random phase due to fluctuating optical path lengths in the experiment. Constructive interference resulted in an enhanced mapping to the spinwave while destructive interference prevented absorption of the input pulses. Examples of constructive and destructive interference are shown in Fig. \ref{fig:traces}. Visibilities were calculated by comparing the most constructive and destructive interference into the spinwave relative to the total optical energy of the input pulses. Visibilities of 70\% were achieved, demonstrating the spatio-temporal matching of the forward and backward pairs. 

We measured memory efficiencies up to $72\pm 5$\% at an optical depth of 500$\pm 100$. The large uncertainty in the efficiency is due to uncertainty in the calibration of detectors and the method for calculating efficiency. The maximum efficiency was calculated by collecting all 2000 pulses and performing a statistical analysis of the results. A statistical distribution was extracted by assuming each data point corresponded to the input pulses from each end generating spinwaves interfering at a random phase $\theta$, with Gaussian noise added to the pulse amplitudes. This produced the distribution for the normalised output pulse energy $\lambda(a+b\sin\theta)$ with $\lambda$ taken from a Gaussian distribution and the random global phase $\theta~\epsilon~[0,2\pi)$. The efficiency corresponds to $a+b$.

While this efficiency is well below the optimal efficiency for this optical depth, and below the demonstrated efficiency for multimode schemes in similar ensembles \cite{cho_highly_2016,hsiao_highly_2018}, it does compare favourably to a Raman memory with a backward retrieval in a similar ensemble \cite{vernaz-gris_high-performance_2018}, which demonstrated efficiency of $65\pm6\%$. To deterministically achieve constructive interference, in  Appendix \ref{appendixpassive} we propose an experimental setup that passively stabilises the phase.

\begin{figure}
\centering
\includegraphics[width=0.5\textwidth]{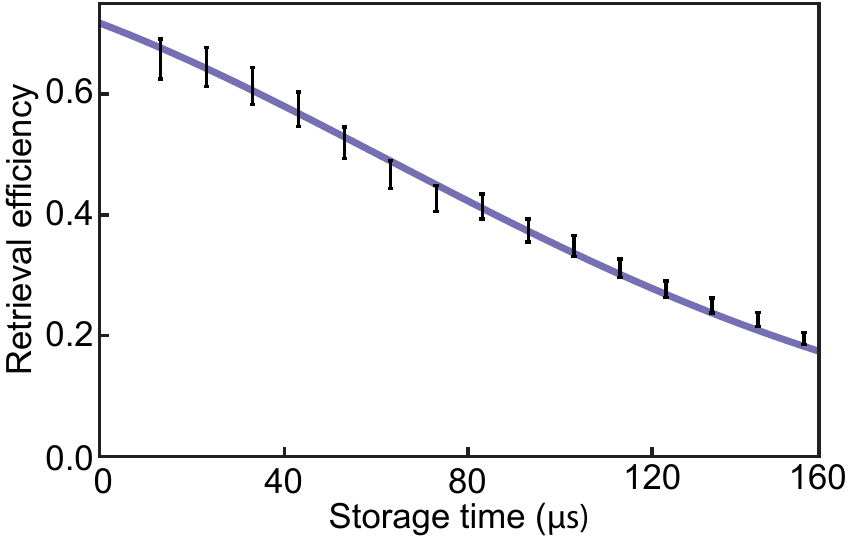}
\caption{Efficiency as a function of storage time. The solid line is a fit with Gaussian decay multiplied by exponential decay $\eta=\eta_0\exp{(-(t/\tau_e)-(t/\tau_g)^2)}$ with constants  $\tau_g=180\pm 20 \si{\micro\second}$ and $\tau_e=250\pm 10 \si{\micro\second}$ \label{fig:storage}}
\end{figure}

\subsection{Memory lifetime}
Memory lifetime was measured by delaying the retrieval of stored pulses, with efficiencies shown in Fig. \ref{fig:storage}. Two decay mechanisms were expected to be important. Due to the phase-matching angle, the phase of the spinwave varies rapidly along the spinwave momentum vector. Motion of the atoms in this direction washes out the spinwave, resulting in a Gaussian decay determined by the temperature of the atoms.  The angle between control and probe due to phase-matching produces a spinwave wavelength of  132 \si{\micro\meter}, much shorter than the wavelength of 4.4 \si{\centi\meter} for exactly co-propagating fields with a 6.83 GHz splitting. At this shorter wavelength we assume the Gaussian component of decay is dominated by atomic motion in this dimension, with the time constant of $(180\pm 20)~\si{\micro\second}$ corresponding to a temperature of $(180\pm20)~\si{\micro\kelvin}$. This is consistent with previous memory experiments in a similar setup \cite{cho_highly_2016}. There is also a clear exponential component in the decay, most likely corresponding to a second decay mechanism of dephasing due to transient magnetic field gradients.

\subsection{The effect of phase-matching}
In a second experiment, we investigated the effects of imperfect phase-matching. As the phase of the spinwave generated varies across the memory and the two inputs interact differently with the spinwave, the single-mode description breaks down. The spatial variation in the memory depends on the relative phase between the two inputs, introducing a difference in the global phase at which each input is maximally absorbed. Without perfect interference between the two inputs, the maximum efficiency is also limited. We exploring this effect by determining and varying the global phase and measuring the efficiency of memory storage and recall. 

\begin{figure*}[h!]
\centering
\includegraphics[width=\textwidth]{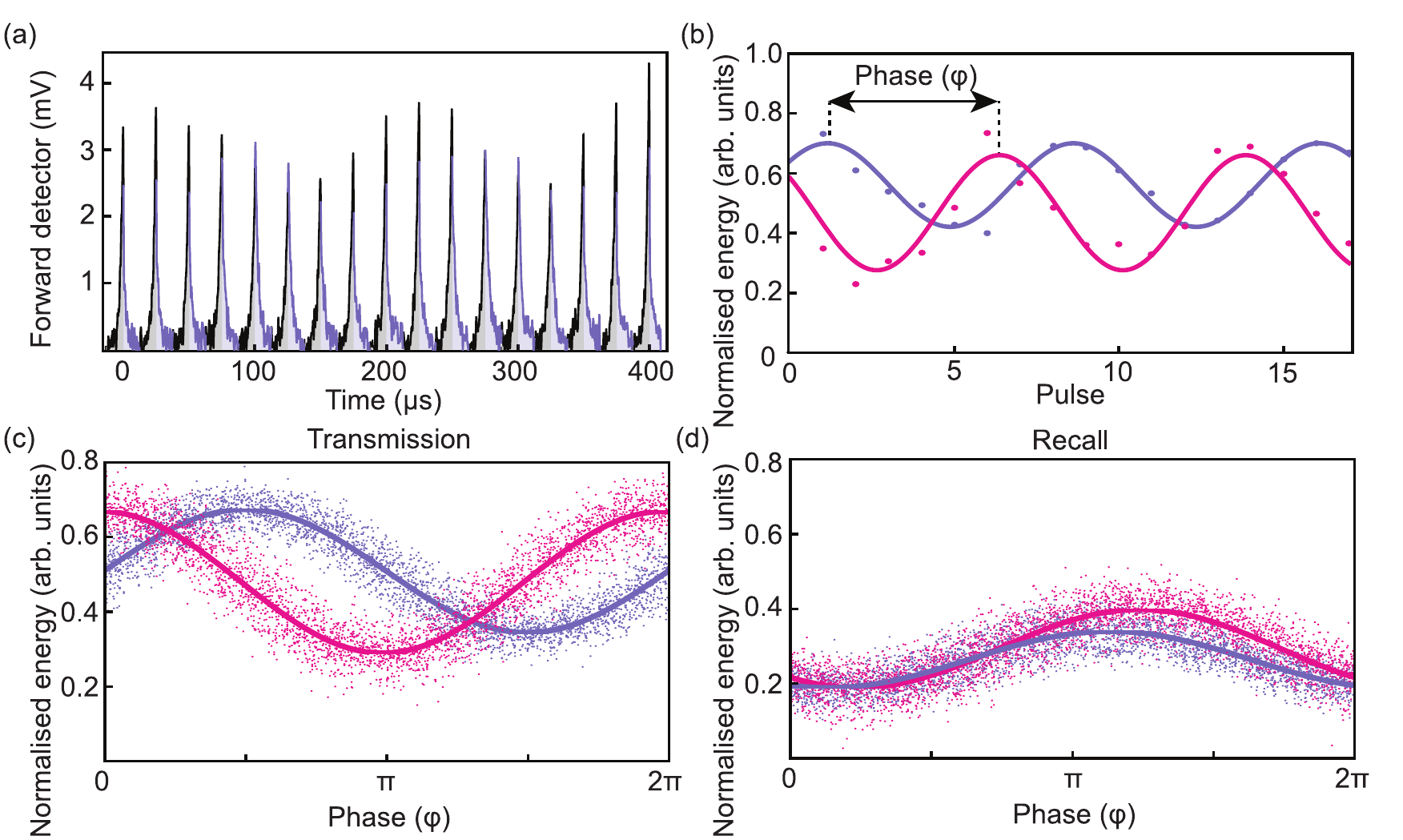}
\caption{(a) Forward detector measurement of a series of sent pulses, with transmitted pulse (black) and recalled pulse (blue). (b) Normalised transmitted energy at forward (blue) and backward (magenta) detectors. Sinusoidal fits allow extraction of the phase $\phi$ between the forward transmitted pulse (blue) and other pulses, in this case the backward transmitted pulse (magenta) (c) All forward (blue) and backward (magenta) pulses transmitted to the detectors during storage, positioned according to the measured phase. Sinusoidal fits allow extraction of visibility and average phase difference. (d) All recalled forward (blue) and backward (magenta) pulses are plotted by phase.\label{fig:phasematch}}
\end{figure*}

As in the first experiment, the global phase $\theta$ between control-probe pairs was random after each loading sequence due to changing optical path lengths, with the system acting similarly to an unlocked interferometer. This phase $\theta$ was stable within each loading phase, and over a period of $400\ \si{\micro\second}$ 17 pulses were stored and retrieved. The dependence of the efficiency on the interference between the pulses due to the phase was tested by incrementing the phase of each successive forward input pulse by $0.3\pi$. This generated over two complete interference fringes per run with a total of $4.8\pi$ incremented to the phase $\theta$ over the duration of the train of pulses. Fig. \ref{fig:phasematch}. (a) has typical photodetector data from one storage sequence. 

The transmitted input pulses and the recalled pulses were all integrated separately and sinusoids were fit for each set of pulses. A typical fitting for the two transmitted inputs is shown in Fig. \ref{fig:phasematch}. (b). The fringe for the forward transmitted pulse is used as a reference and the phase differences of the fringes for the other integrated pulses are used to determine the quality of the phase matching. Ideally, there would be no phase difference between the fringes for the transmitted inputs, or between the two recalls, and a $\pi$ phase difference between transmission and recall.

In Fig. \ref{fig:phasematch}. (c) and (d) the plotted fringes show the average phase differences. Each individual pulse is also added to the plot to demonstrate the variance in the data. The phase difference between the transmitted inputs in (c) is $1.6\pm0.2$ rad. The phase difference between the recalled outputs was smaller at $0.4\pm0.2$ rad. These measurements agree with theoretical predictions which are discussed in Appendix \ref{appendixphase}. Three datasets were taken, adjusting the phase-matching by alignment of the control fields. The smallest measured phase difference between inputs was $0.5\pm0.2$ rad.

\section{Conclusion}
In summary, we have demonstrated a novel optical memory based on a distributed reflection within an atomic ensemble. This scheme has the improved efficiency expected of a single-mode memory without any need of an optical cavity. This may be advantageous for applications based on free-space coupling of light to atomic ensembles.
\begin{acknowledgments}
This research was conducted by the Australian Research Council Centres of Excellence Centre for Quantum Computation and Communication Technology (CE110001027).
\end{acknowledgments}

JLE designed the experiment. JLE, PVG, ADT, KP, AL, and GTC performed the experiment. JLE and PVG analysed the data, and JLE and KP performed numerical calculations. All authors contributed to the manuscript.

Correspondence should be directed to BCB (ben.buchler@anu.edu.au)

\appendix
\section{Efficiency of TRACE memory \label{appendixeffic}}
A theoretical maximum efficiency can be calculated following the method of \cite{gorshkov_photon_2007-1}. The fundamental efficiency limit in that theory comes from the decay of the excited state coherence $P$. We calculate the efficiency of a complete retrieval of the spinwave from the memory, and detail the assumptions necessary to perform a time-reversal of this process. Storage corresponds to the same process, time-reversed, and therefore yields the same efficiency.

To measure the retrieval, we assume the memory is initialised with $|S|\neq 0$ and set the input $\mathcal{E}_{IN}=0$,

\begin{align}
\mathcal{E}_{OUT\pm} &=\mathcal{E}_{IN\pm}+i\sqrt{d}P_\pm \label{eq:suppefficiency1}\\
\partial_t P_+ &= -(\Gamma + i\Delta)P_+ + i\Omega S + i \sqrt{d}\Gamma\mathcal{E}_{OUT+} \nonumber\\
 &= -\left(\Gamma(1+dz) + i\Delta\right)P_+ +i\Omega S \label{eq:suppefficiency2}\\
\partial_t P_- &= -\left(\Gamma(1+d(1-z)) + i\Delta\right)P_- +i\Omega S\\
\partial_t S &= i \Omega^*(P_++P_-).\label{eq:suppefficiency3}
\end{align}

Combining equations \eqref{eq:suppefficiency1}, \eqref{eq:suppefficiency2}, and \eqref{eq:suppefficiency3} leads to the complete time derivative of the spinwave, 
\begin{align}
d/dt(|S|^2+|P_+|^2+|P_-|^2) &=-2\Gamma(|P_+|^2+|P_-|^2)+2d\Gamma(z|P_+|^2+(1-z)|P_-|^2) 
\end{align}

Now in the adiabatic limit, $P_+=P_-$. Substituting and integrating over all time gives
\begin{align}
\int_0^\infty|\mathcal{E}_{OUT}|^2\mathrm{d}t&=\frac{d}{d+2}\left[|S(0)|^2+|P_+(0)|^2+|P_-(0)|^2-|S(\infty)|^2-|P_+(\infty)|^2-|P_-(\infty)|^2\right]
\end{align}
which results in a storage or retrieval efficiency of $\frac{d}{d+2}$ and $\left(\frac{d}{d+2}\right)^2$ for storage followed by retrieval.

\section{Phase-matching \label{appendixphase}}
\begin{figure}[htbp!]
\centering
\includegraphics[width=0.5\textwidth]{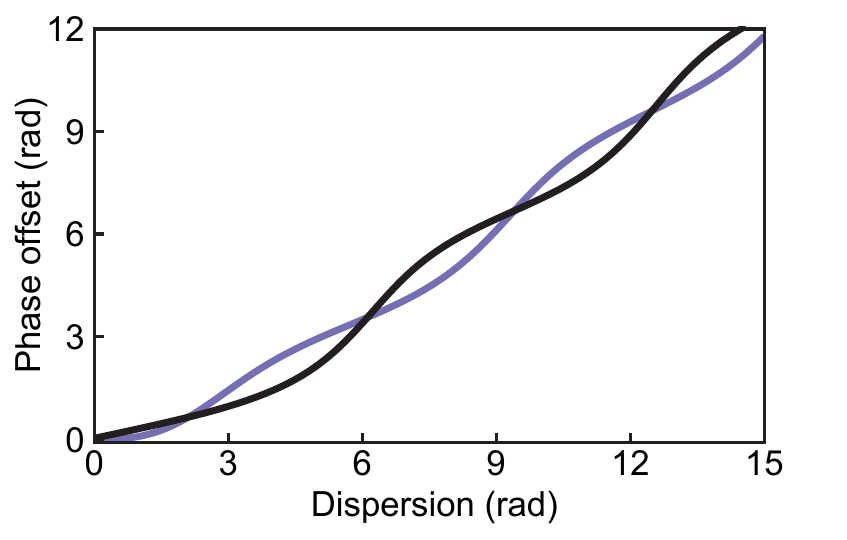}
\caption{Phase offsets between forward and backward optical maxima for transmitted light (black) and recalled light (blue) for the total unmatched dispersion of each field crossing the memory. The vertical axis is continued beyond $2\pi$ in the vertical to show the continuous dependence of the phase offset. \label{fig:dispersion}}
\end{figure}

Due to the existence of additional transitions, the dispersion is not entirely matched under the detuning $\Delta_{p-}=-\Delta_{p+}$. On the rubidium-87 D1 line the  additional transition to $\ket{5P_{1/2}, F=2}$ is far detuned from both probes under our experimental conditions, and the phase mismatch can be accounted for by small adjustments to one or both single-photon detunings. Without those adjustments a phase offset of 0.14 rad is expected for an optical depth of 500 and a detuning from $\ket{5P_{1/2}, F=1}$ of $\pm 230~\si{\mega\hertz}$. Fig. \ref{fig:dispersion} shows the expected phase offset for inputs and outputs for a given unmatched dispersion across the memory, under typical experimental conditions.

Sufficiently bad phase-matching separates the scheme into two uncoupled memories; the rapidly varying spatial phase of a spinwave recalled in the opposite direction prevents the driving of that spinwave by the opposite control field. The main cause of poor phase-matching in the experiment was determined to be alignment, and the effects of poor phase-matching can be seen in Fig. \ref{fig:phasematch}.

Finally, the presence of multiple excited states with appropriate coupling strengths can allow for frequencies at which no dispersion is applied to the optical field. Then, both input fields can be given the same frequency, and similarly the control fields. Where the two input fields have equal frequencies, the effects of the standing wave generated in the control field must be suppressed.

\section{Passive global phase stabilisation \label{appendixpassive}}
\begin{figure}[htbp!]
\centering
\includegraphics[width=0.5\textwidth]{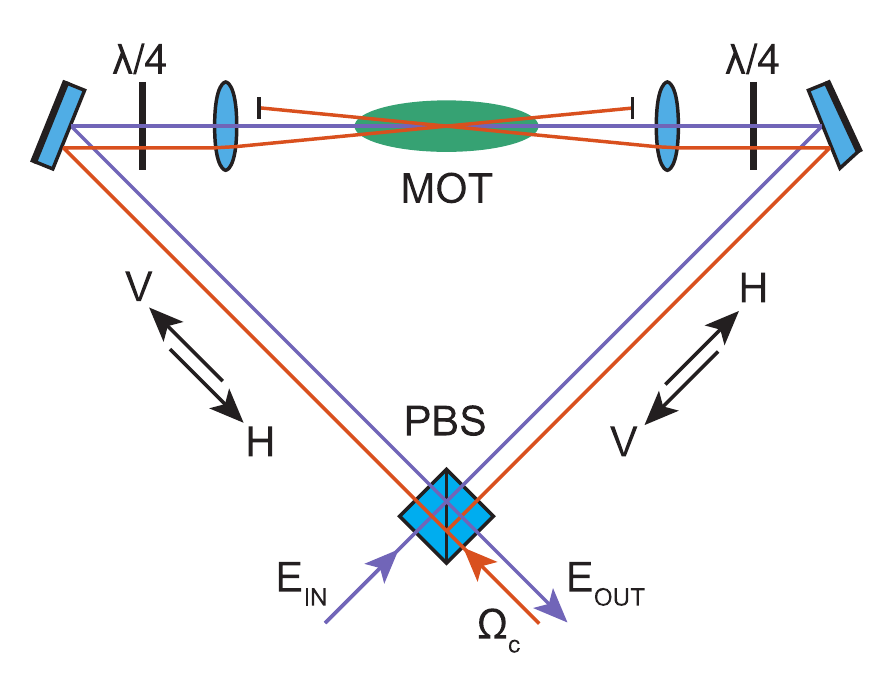}
\caption{Sagnac configuration for stable TRACE memory. A diagonally polarised input is split by the PBS and sent to the memory. The quarter wave plates rotate the polarisations of the two inputs such that they interact with the same atomic transition. The second rotation sends the recalled pulses through the output port of the PBS. The common path shared by the control-probe pairs protects against optical path length changes. \label{fig:sagnac}}
\end{figure}

If a single frequency for both inputs can be used, the Sagnac configuration shown in Fig. \ref{fig:sagnac} would allow for a passive stabilisation of the global phase. Each control-probe pair shares an optical path, and so minimal variation in their relative phase would occur. The phase between the pairs can be adjusted by coarse path length changes, or by changing the ellipticity of the polarisation of the input or control fields. The output is also sensitive only to large changes in optical path length.

\bibliographystyle{apsrev4-1}
\bibliography{trace}
\end{document}